# Analysis of the result of the Neutrino-4 experiment in conjunction with other experiments on the search for sterile neutrinos within the framework of the 3 + 1 neutrino model

A.P. Serebrov, R.M. Samoilov, M.E. Chaikovskii

Petersburg Nuclear Physics Institute, National Research Center Kurchatov Institute, 188300, Gatchina, Russia

serebrov_ap@pnpi.nrcki.ru

Abstract

The correspondence of the results obtained in the Neutrino-4 experiment with the results of the NEOS, DANSS, STEREO, PROSPECT experiments at reactors, the MiniBooNE, LSND, MicroBoone experiments at accelerators, the IceCube experiment and the BEST experiment with a $^{51}$Cr neutrino source is analyzed. The agreement between the results of the Neutrino-4 experiment, the BEST experiment and the gallium anomaly on mixing angel is discussed. The disagreement between the results of the above-mentioned direct experiments with the results of the reactor anomaly, as well as with the limitations from solar and cosmological data, is discussed. It is shown that the results of the above-mentioned direct experiments on the search for sterile neutrinos and IceCube experiment do not contradict the Neutrino-4 experiment within the framework of the 3+1 neutrino model at the available experimental accuracy. The sterile neutrino parameters from the Neutrino-4 and BEST experiments make it possible to estimate the sterile neutrino mass $m_4 = (2.70 \pm 0.22)$eV and the effective mass of the electron neutrino $m_{4\nu_e} = (0.82 \pm 0.21)$eV. The matrix with absolute values of the 3 + 1 neutrino model mixing parameters and the mixing scheme are presented.

Keywords: sterile neutrino, neutrino mass, neutrino mixing.

## 1. Introduction

The Standard Model includes three massless neutrinos that participate in weak interactions. Measurement of the total decay cross section of a neutral Z-boson imposes a limitation on the number of active neutrinos. At the moment, this limitation is $N_\nu = 2.92 \pm 0.05$ from the direct measurement of invisible Z width ($N_\nu = 2.996 \pm 0.007$ from Standard Model fits to LEP-SLC data) [1]. Experimentally observed neutrino oscillations require the introduction of nonzero neutrino masses and the mixing matrix. The mixing parameters of the three flavor states of the Standard Model are determined experimentally. However, there are also experimentally observed anomalies that cannot be described within the framework of the three neutrino mixing. The anomalies have been observed in several accelerator and reactor experiments: LSND at the CL 3.8 σ [2], MiniBooNE 4.7 σ [3], the Reactor antineutrino anomaly (RAA) 3σ [4,5], and in the experiments with the radioactive sources GALLEX/GNO and SAGE 3.2σ – gallium anomaly (GA) [6-8]. There is a direct way to expand the theory to explain these phenomena - adding sterile neutrinos to the theory of particles. One of the possible extensions of the theory is the 3 + 1 model with one sterile state and one additional mass state of the order of several eV. Consideration such a neutrino as a possible particle of minor component of the dark matter is limited by cosmological constraints [9], which can be overcome by extension of the cosmological model [10].

Earlier, a comparison of the results of the Neutrino-4 experiment and the results of the above works was carried out in the work [11] as a part of data analysis of the Neutrino-4 experiment. This work is devoted to discussing the possibility of experimental confirmation of the 3 + 1 model based on the latest data from the Neutrino-4 experiment, which were recently published in the work [12].

## 2. Results of the Neutrino-4 experiment

The Neutrino-4 experiment [12] is carried out at the SM-3 reactor (Dimitrovgrad, Russia). Reactor neutrino experiments, in which the measured neutrino flux is compared with the calculated theoretical value, have a serious drawback - they depend on the accuracy of the expected flux calculations and detailed information about the detector efficiency. However, in the Neutrino-4 experiment, due to the small size of the active zone (35x42x42cm$^3$) of the SM-3 reactor and the high power of 90 MW, it is possible to apply the method of relative measurements using a movable detector. In the mixing model with one sterile neutrino, the flux of reactor antineutrinos at small distances depends on the ratio of distance to energy according to expression (1).

$$P(\bar{\nu}_e \to \bar{\nu}_e) = 1 - \sin^2 2\theta_{14} \sin^2\left(1.27\frac{\Delta m_{14}^2[eV^2]L[m]}{E_{\bar{\nu}}[MeV]}\right) \quad (1),$$

where $E_{\bar{\nu}}$ antineutrino energy in MeV, L is distance in meters, $\Delta m_{14}^2$ is the – the difference between the squares of the masses of the first and fourth neutrino mass states, $\theta_{14}$ is the angle parameterizing the mixing matrix and responsible for mixing the electron neutrino with the fourth mass state. Experimental verification of oscillations caused by mixing with a state having a mass of the order of several eV requires measurements of the neutrino flux at as small distances as possible from an almost point like antineutrino source. A detailed description of the Neutrino-4 detector, preparation of the neutrino laboratory, background measurements and the scheme of the data acquisition process are presented in [12]. Here we briefly outline the method and results of direct observations of oscillations in the Neutrino-4 experiment.

Antineutrino flux measurements are carried out by comparing the detector signals measured with the operating and stopped reactor. The difference between these signals is the measured antineutrino flux, while the background of cosmic rays, which is the main problem for the measurement, is cancelled out. We use the method of relative measurements, because the same detector measures at different distances.

The measured flux as a function of energy and distance to the reactor is conveniently represented in the form of a matrix $N(E_i, L_k)$ [12]. The well-known problem of the difference between the calculated and experimentally observed spectra of reactor antineutrinos was also observed in our experiment [12]. The existence of this problem makes it necessary to apply the analysis method which reduces the influence of the calculated spectrum. In order to use the method of relative measurements, we compose the ratio according to the equation (2). The numerator is the measured number of neutrino events multiplied by the square of the distance, and the denominator is the expected value for the same energy averaged over the entire distance range. The advantage of this expression is that the dependence on the energy spectrum is reduced as well as the dependence on the efficiency of the detector.

$$R_{ik}^{\text{exp}} = (N_{ik} \pm \Delta N_{ik})L_k^2 / K^{-1} \sum_k^K (N_{ik} \pm \Delta N_{ik})L_k^2 = \frac{\left\langle S(E)\left(1 - \sin^2 2\theta_{14} \sin^2\left(\frac{1.27 \Delta m_{14}^2 L_k}{E}\right)\right)\right\rangle_i}{K^{-1} \sum_k^K \left\langle S(E)\left(1 - \sin^2 2\theta_{14} \sin^2\left(\frac{1.27 \Delta m_{14}^2 L_k}{E}\right)\right)\right\rangle_i} \quad (2)$$

S(E) is initial $^{235}$U antineutrino spectrum, $\langle s \rangle$ means integration with function of energy resolution with σ=250 keV and integration over energy bin intervals. If the summation is performed over distances that significantly exceed the length of the oscillations, then the denominator is simplified:

$$R_{ik}^{\text{th}} \approx \frac{1 - \sin^2 2\theta_{14} \sin^2(1.27 \Delta m_{14}^2 L_k / E_i)}{1 - \frac{1}{2}\sin^2 2\theta_{14}} \xrightarrow[\theta_{14}=0]{} 1 \quad (3)$$

Equation (3) characterizing the observed dependence differs from equation (1) only by a constant factor $1 - 1/2 \sin^2 2\theta_{14}$ in the denominator. Consequently, the process of oscillations can be directly obtained from the experimental data if the result is presented in the form of the L/E dependence. Thus, we fixate the phase of the oscillation process for different energies. The result of data analysis is shown in FIG.1a. The period of oscillations for an average antineutrino energy 4 MeV is 1.4 m, that is, less than the biological shielding of a reactor - 5 m. Despite this, the oscillation process can be observed provided that the method described by us is used. This method of a fixed phase, like the holographic method, allows one to restore the image.

Since we observe the effect of oscillations, it is important to check that this cannot be a false instrumental effect. The effect disappears when the reactor is stopped. The bottom part of FIG.1a shows the result of processing the cosmic background data using the method applied to the analysis of the antineutrino flux. The fast neutrons of the cosmic background give a correlated signal similar to the correlated signal of the antineutrino events. The recoil proton from the fast neutron scattering at the proton produces the prompt signal which imitates the positron signal, and then the neutron is captured by gadolinium, producing the delayed signal. Thus, the entire process of signal registration and data processing can be monitored by measuring the signals of the correlated background. The result of the analysis of the cosmic background is shown in the bottom part of the FIG. 1a. The correlated background generated by fast neutrons has a slight dependence on distance, since there is an inhomogeneity in the passive shielding of the detector by concrete in the building structure. In calculating the R matrix for the neutrino signal, we introduced a correction for the distance, and for calculating the background, we introduce a correction for this inhomogeneity of the cosmic background. The dependence of the background R matrix on the L / E ratio is consistent with the hypothesis of the absence of oscillations with $\chi 2 \ / \ DoF = 1.3$, while the disagreement with the oscillation curve is at the level of $\chi 2 \ / \ DoF = 6.1$. This means that the observed oscillations are not a systematic instrumental effect. The decay of the oscillation curve is related to the finite energy resolution of the detector.

Finally, it is important to note that the oscillation curves have the first maximum, which lies at the point L = 0, since the process of oscillations starts from the neutrino source. FIG. 2 shows the oscillation curve completed to the source, which confirms that the source of the oscillations is the reactor.

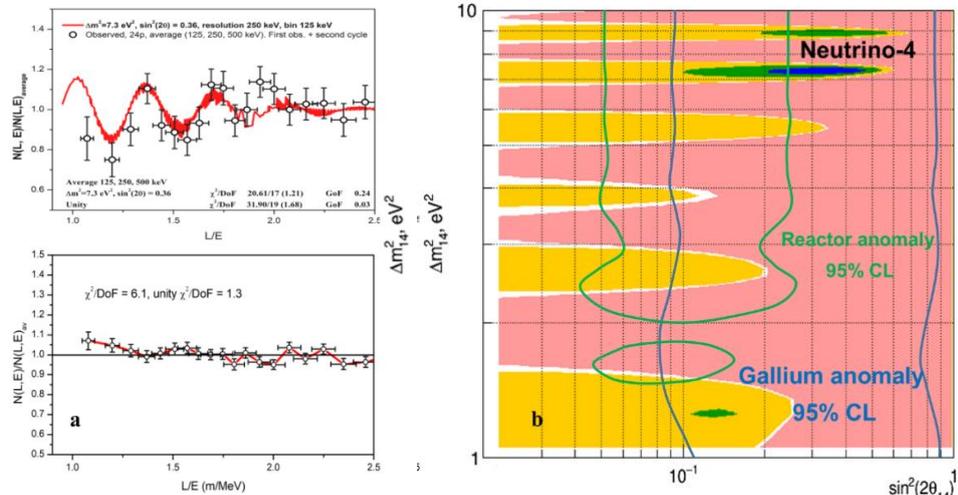

FIG.1 a - Oscillatory curve of the neutrino signal (reactor ON minus reactor OFF) and data processing by the same method with the stopped reactor, demonstrating the absence of a systematic effect. b - Results of data analysis on the plane $\Delta m_{14}^2, \sin^2 2\theta_{14}$.

Comparison of the experimental matrix and the matrix obtained as a result of MC simulation can also be performed by the $\Delta\chi^2$ method, that is, using the expression: $\sum_{i,k}(R_{ik}^{exp} - R_{ik}^{th})^2/(\Delta R_{ik}^{exp})^2 = \chi^2(\sin^2 2\theta_{14}, \Delta m_{14}^2)$. The result of the analysis by the $\Delta\chi^2$ method is shown in FIG. 1b. The area of the parameter, colored in pink, is excluded with a CL of 3σ. However, in the region $\Delta m_{14}^2 = 7.3 eV^2$ and $\sin^2 2\theta_{14} = 0.36 \pm 0.12_{stat}$ the oscillation effect is observed at the CL of 2.9 σ, Monte-Carlo based analysis shows 2.7σ CL [12]. For comparison, the parameter regions of the gallium and reactor anomalies are presented together with the Neutrino-4 results.

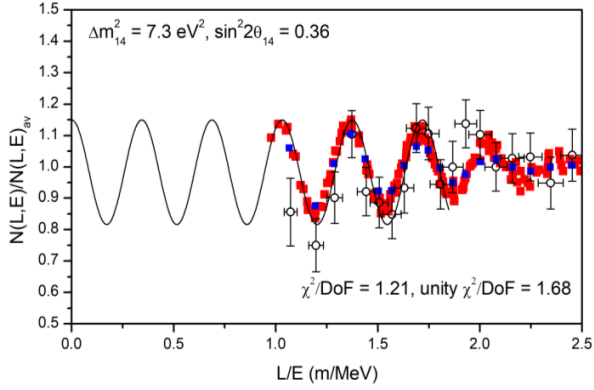

FIG.2. Full curve of the oscillation process from the center of the reactor core.

### 3. Comparison of the results of the experiment Neutrino-4 with gallium anomaly (GA), reactor antineutrino anomaly (RAA) and the solar model.

The Neutrino-4 experiment is aimed at direct measurement of the oscillation parameter $\sin^2 2\theta_{14}$, the value of which is twice the deficit of the total neutrino flux at distances significantly larger than the oscillation length. Therefore, to compare the results of Neutrino-4 with experiments measuring the total neutrino flux (RAA and GA), the value $\sin^2 2\theta_{14}$ can be used to calculate the deficit, or vice versa. We compare the results in terms of the oscillation parameter $\sin^2 2\theta_{14}$.

Oscillation parameters values (best fit point) in gallium experiments considering result of the BEST experiment is $\sin^2 2\theta_{14} = 0.34^{+0.14}_{-0.09}, \Delta m_{14}^2 = 1.25^{+\infty}_{-0.25}эВ^2$ [13]. This result is in good agreement with the result of the Neutrino-4 experiment: $\sin^2 2\theta_{14} \approx 0.36 \pm 0.12(2.9\sigma)$ in $\Delta m_{14}^2 > 5 eV^2$ area.

The BEST experiment was designed to find oscillation parameters $\sin^2 2\theta_{14}$ and $\Delta m_{14}^2$ and has a high sensitivity at $\Delta m_{14}^2 \sim 1 eV^2$ area. So, with this purpose detector was divided into two volumes. But with value of $\Delta m_{14}^2 \sim 7 eV^2$ oscillations are averaged in both volumes, so there are difficulties in determining $\Delta m_{14}^2 > 5 eV^2$. However, in the experiment the deficit of the neutrino flux for both volumes was measured with sufficiently high accuracy ($R_{inn} = 0.79 \pm 0.05$ and $R_{out} = 0.77 \pm 0.05$). A detailed analysis of attempts to determine $\Delta m_{14}^2$ is made in [13]. For the mixing angle parameter, the value given in [13] is $\sin^2 2\theta_{14} = 0.42^{+0.15}_{-0.17}$ for the BEST experiment only and $\sin^2 2\theta_{14} = 0.34^{+0.14}_{-0.09}$ (for $\Delta m_{14}^2 > 1 eV^2$) for the BEST experiment jointly with the gallium anomaly (GALLEX and SAGE). The simplest joint estimate for $\sin^2 2\theta_{14}$ for GA and BEST with Neutrino-4 can be done can be done under the assumption that the effect is in the region of fast oscillations, as indicated by the Neutrino-4 experiment. Then the average value $\sin^2 2\theta_{14}$ from the result of Neutrino-4 and GA is $\sin^2 2\theta_{14} = 0.35^{+0.09}_{-0.07}$. Thus, the CL of the presence of oscillations is 5σ. Comparison of the result of the Neutrino-4 experiment and the result of the BEST experiment is shown in Fig. 3 on the left. An analysis was also performed based on the GALLEX, SAGE and BEST experiments data published in [13]. $\chi^2(\Delta m_{14}^2, \sin^2 2\theta_{14})$ distribution was reproduced. Using this result together with the distribution from the Neutrino-4 experiment, the distribution $\Delta\chi^2(\Delta m_{14}^2, \sin^2 2\theta_{14})$ was obtained and presented in Figure 3 on the right. The parameters value at the best fit point are $\sin^2 2\theta_{14} = 0.38$, $\Delta m_{14}^2 = 7.3 эВ^2$. The confidence level of the observation of oscillations is 5.8σ

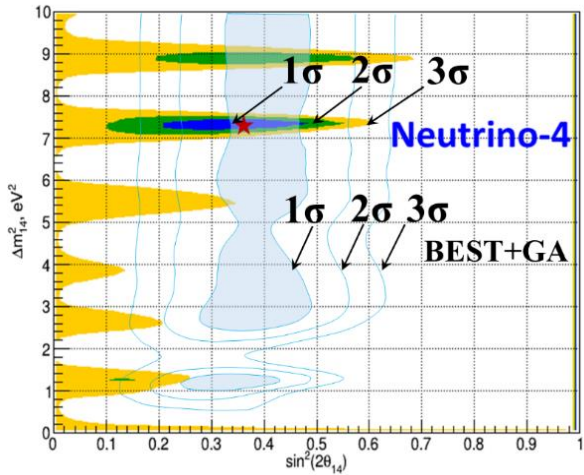 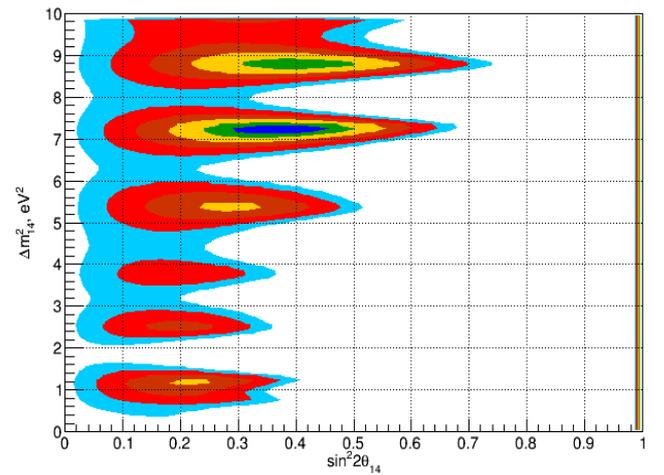

FIG. 3. On the left – comparison of the result of the BEST experiment with GA and the result of the Neutrino-4 experiment. On the right – The result of the combined analysis of GA, BEST and Neutrino-4, where blue indicates the area with a 1σ CL, green - 2σ, yellow - 3σ, dark red - 4σ, red - 5σ and blue - 5.8σ.

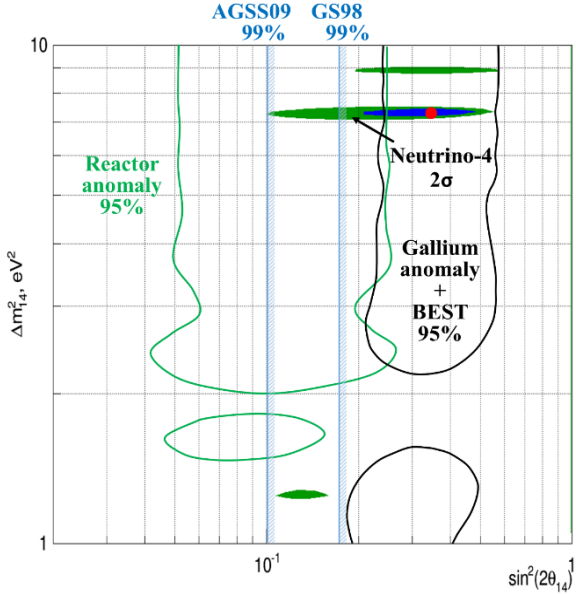

FIG.4. Comparison of the results of the Neutrino-4 and GA with BEST experiments with the reactor antineutrino anomaly [15] and the solar model [16].

Comparison of the experimental effect in RAA, GA and Neutrino-4 for the mixing angle can be done, as before, under the assumption of fast oscillations with the parameter $\Delta m_{14}^2 \sim 7$ eV$^2$. Then, the deficit of the RAA according to the Huber-Muller model is $R = 0.930^{+0.024}_{-0.023}$ [14] and estimation for $\sin^2 2\theta_{14}$ is $\sin^2 2\theta_{14} \approx 0.13 \pm 0.04$. Comparing this estimate with the estimate $\sin^2 2\theta_{14} \approx 0.35^{+0.09}_{-0.07}$ shows discrepancy which is 2.4σ. Although this discrepancy has not yet gone beyond 3σ, a possible interpretation of this situation is required. The limitations of PAA [15] and GA together with BEST are shown in FIG 4.

The interpretation of the RAA measurements is based on the calculated spectrum, which still has unexplained discrepancies with the measured spectrum. Thus, calculations show that there are discrepancies in the integrals of the spectra and in their shapes. There is a so-called "bump". If there is an excess in the spectrum in the 5 MeV region, can be assumed some isotopes and decays are not taken into account. But first, one should take into account the oscillation effect found in the Neutrino4 and gallium experiments. FIG. 5 (left) shows the antineutrino spectra for $^{235}$U and $^{239}$Pu from [17] in comparison with the calculated spectra according to the Huber-Mueller (HM) model. If we take into account the effect of oscillations, with the parameter $\sin^2 2\theta_{14} \approx 0.35^{+0.09}_{-0.07}$, then we can obtain the result of a comparison of the spectra shown in FIG. 5 (right). Of course, there is still a "bump" in the 5-6 MeV region in relation of the experimental to the theoretical spectrum, so it is necessary to look for unaccounted nuclides. It should be noted that the difficulty of calculations is characterized by the following set of results for the R-ratio of observed to expected flux: $0.930^{+0.024}_{-0.023}$ (2.8σ), $0.975^{+0.032}_{-0.030}$ (0.8σ), $0.922^{+0.024}_{-0.023}$ (3.0σ), $0.970 \pm 0.02$ (1.4σ), $0.960^{+0.022}_{-0.021}$ (1.8σ) [14]. Thus, the accuracy of the calculations does not allow making a specific conclusion about the size of the anomaly, that is, about the size of the deficit. For determining the deficit from the effect of oscillations in the Neutrino-4 experiment, we can use that for $\Delta m_{14}^2 = 7.3$ eV$^2$, the effect of averaging oscillations at distances greater than 10 meters is realized. Therefore, the ratio $R = 1 - \frac{1}{2}\sin^2 2\theta$ will be valid and predicts $R = (1 - 0.18) \pm 0.06 = 0.82 \pm 0.06$. At the same time, all the calculated values of the R-ratio listed above turn out to be higher than the expected value from the oscillation effect.

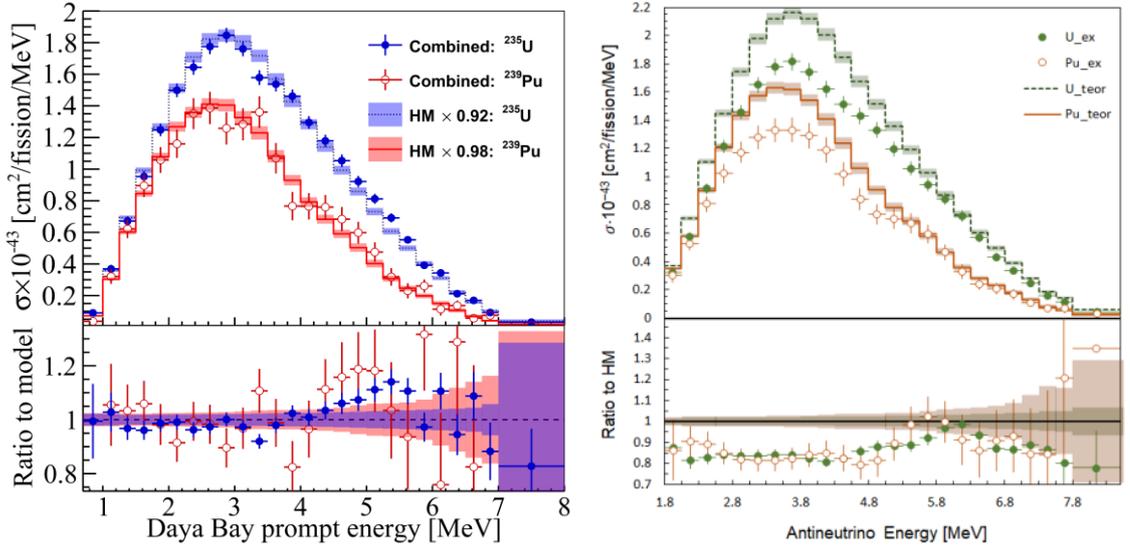

FIG. 5. Left - antineutrino spectra for 235U and 239 Pu from work [17] in comparison with the calculated spectra according to the Huber-Mueller (HM) model, right –comparison of the spectra if the parameter $\sin^2 2\theta_{14} \approx 0.35^{+0.09}_{-0.07}$ is taken into consideration.

FIG. 6 shows the cumulative yield of nuclides upon fission of $^{235}$U as a function of the decay energy and half-life. The ENDF/ B-VII.1 database was used to construct the distribution. This distribution contains 693 nuclides. Nuclides with high energy decay within seconds and fractions of a second. The difficulties in experimentally measuring the schemes of such fast decays are obvious. The region with an energy above 4 MeV and a half-life of no more than 10 seconds is 14.4. The excess of the spectrum of this region of the so-called "bump" is approximately 15%, so the proportion of isotopes which were not considered is $0.144 \cdot 0.15 = 0.02$, only 2%.

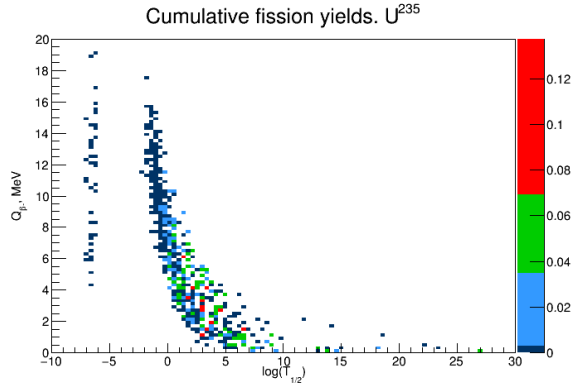

FIG.6. Cumulative yield of nuclides during fission of $^{235}$U depending on the decay energy and half-life.

In principle, the question should be asked which results should be preferred. Results of direct experiments or results obtained in the course of calculating complex processes? Based on the above consideration, the answer is obvious. It should be noted that the RAA is based on a rather complicated method of absolute measurements, but the Neutrino-4 experiment uses the method of relative measurements and does not require exact knowledge of the spectrum of reactor antineutrinos. The BEST experiment uses the well-known spectrum of monochromatic neutrinos and is also more reliable. It uses a method successfully applied earlier in the SAGE experiment [7].

We can say that the discrepancy between the result of Neutrino-4 and RAA is the discrepancy between direct and indirect, depending on complex calculations, measurements.

The second contradiction between the results of the Neutrino-4 and BEST experiments with GA is observed with restrictions on $\sin^2 2\theta_{14}$, based on the solar neutrino model (fig.4). This result was also obtained in the course of calculating complex processes, where there are the following problems.

1. Interpretation of measurements of the solar neutrino flux is based on the theoretically calculated value of the total solar neutrino flux. This value is calculated on the basis of the Standard Solar Model, and therefore includes the uncertainties of this model and the unresolved problem of the solar metallicity.
2. The probability of detecting an electron neutrino in the total flux of solar neutrinos depends on all mixing angles. This means that the accuracy of the calculations is determined by the accuracy of knowing the mixing angles.
3. For neutrinos produced during the decay of $^8$B, the theoretical flux includes taking into account the MSW effect, which is calculated using the electron density distribution in the Sun and the adiabatic approximation.

Recent work describing restrictions on $\sin^2 2\theta_{14}$ based on global analysis of experimental data uses primarily the data of Borexino, SK and SNO [16]. The restrictions on $\sin^2 2\theta_{14}$ in the two metallicity models differ by a factor of 1.8. This indicates that the systematic uncertainties in neutrino calculations based on the solar model have not yet been determined with sufficient accuracy. The accuracy of calculation of the neutrino flux from the main pp chain in the SSM is 0.6%, but the accuracy of the experiment is 16%. The boron neutrino yield is a millionth of the total neutrino yield, which means that the accuracy of model calculations is extremely important. The boron neutrino yield was measured with an accuracy of 4%, if we add up the results of all experiments, although the joint inclusion of systematic errors in experiments leaves questions. The uncertainty in the calculation of the boron neutrino yield is ±12%, and the calculation results for the two models differ by 17% [16]. There is an experimental accuracy, but there is no calculation accuracy for boron neutrinos, and vice versa, there is a calculation accuracy, but there is no experimental accuracy for a pp chain.

First of all, we estimated (independently of the solar model) the possibility of making a choice between the 3-neutrino model and the 4-neutrino model, using the current accuracy of the mixing matrix elements for the two neutrino models. When calculating the electron neutrino yield in the 3+1 model, one should take into account that the matrix elements U$_1$, U$_2$, U$_3$ must be corrected due to the introduction of a new mixing angle. In this case, the errors of the newly introduced mixing angle must also be taken into account. It can be seen that the transition from considering the problem within the framework of the 3 neutrino model to the 3+1 model significantly decreases the accuracy of the estimate. FIG. 7 shows the estimates (1 σ) for mixing in vacuum for the variant of 3 neutrinos and 4 neutrinos in accordance with the formula: $P_{ee} = \sum_{k=1}^{n}|U_{ek}|^4$. A similar calculation was made for the high-energy part of the spectrum (for boron neutrinos), assuming that the MSW effect leads to the approximation $\sin\theta_{12} = 1$. The FIG.7 demonstrates results of two calculations: with linear and quadratic addition of errors. Since statistical errors add up quadratically, and systematic errors add up linearly, we present two extreme cases for comparison, since strict separation of the contributions of those and other errors of matrix elements is difficult. For the 3-neutrino model and for the 4-neutrino model, we have the following criteria of significance for experimental results in different models. We have $GoF_{3\nu} = 0.895$, $GoF_{4\nu} = 0.09$ for quadratic addition of errors and the confidence levels of agreement with the hypotheses are $0.13\sigma$ and $1.67\sigma$ respectively. This difference does not yet allow one of the hypotheses to be reliably preferred. And we have $GoF_{3\nu} = 0.9$, $GoF_{4\nu} = 0.33$ for linear addition of errors and corresponding confidence levels of agreement are 0.12σ and 0.43σ. In this case, the difference is even smaller, so it is even more impossible to obtain a convincing difference between the hypotheses. The limitations on $\sin^2 2\theta_{14}$, obtained in [16] are beyond the available experimental accuracy of the mixing matrix and the accuracy of neutrino experiments. The claimed contradiction between experiments is heavily based on SSM, but solar models still contain significant uncertainties, while direct measurements of neutrino fluxes from reactors or radioactive sources rely less on calculations. This contradiction, rather, indicates the need to revise the SSM, if the result of Neutrino-4+BEST+GA is confirmed.

**4. Comparison of the results of Neutrino-4 with the results of the PROSPECT, STEREO, DANSS and NEOS experiments**

The comparison of the Neutrino-4 results with the earlier results of other reactor neutrino experiments DANSS [19], NEOS [20], PROSPECT [21], and STEREO [22] can be found in [11]. An illustration of the comparison with reactor experiments is shown in FIG. 8 on the left. The DANSS and NEOS experiments at NPPs have a significantly lower sensitivity to large values of the parameter $\Delta m_{14}^2 > 3$ eV$^2$ due to the large size of the reactor core, 3-4 meters. For oscillations with a period of 1.4 m (for an average antineutrino energy of 4 MeV with $\Delta m_{14}^2 \sim 7$ eV$^2$), the effect is averaged already within the reactor core. Therefore, DANSS and NEOS constraints with 95% CL in FIG. 8(left) do not reach the values $\Delta m_{14}^2 \geq 7$ eV$^2$, and for $\Delta m_{14}^2 \geq 5$ eV$^2$ they cover only large values of $\sin^2 2\theta_{14} \geq 0.5$.

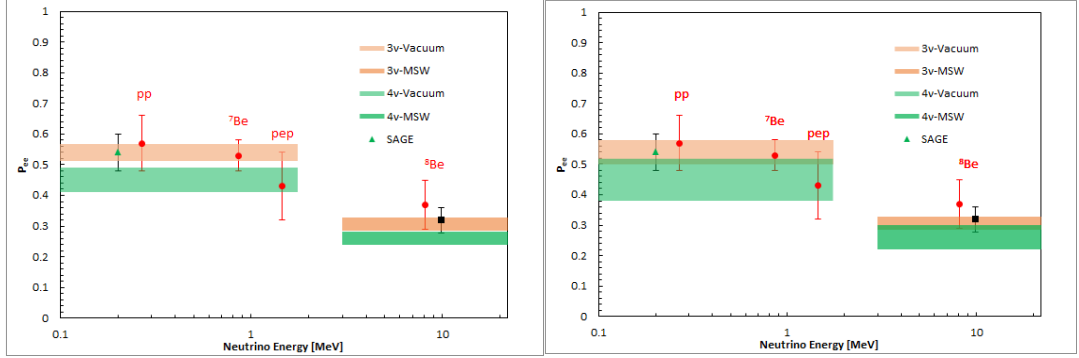

FIG.7. Two calculations with linear and quadratic addition of the errors of the mixing matrix elements for the 3-neutrino model and the 4-neutrino model. In the left figure, all errors are added quadratically. In the right figure, all errors are added linearly. SAGE experiment data are taken from [18].

The result of the PROSPECT experiment [21] does not fully agree with the result of Neutrino-4. The region of 7.3 eV$^2$ with a confidence level of ~1σ Neutrino-4 is overlapped on the PROSPECT 95% CL exclusion area. Although, it should be noted that the exclusion contours obtained in STEREO [22] are quite close to the allowed oscillation parameters from the Neutrino-4 and the best fit point of the STEREO experiment is in the area of the result of the Neutrino- 4. It is especially important that in the Neutrino-4 experiment we observe the oscillatory process directly in the measurements. Probably, with a higher sensitivity in the STEREO experiment will observe the effect of oscillations with parameters close to $\Delta m_{14}^2 \sim 7$ эВ$^2$, $\sin^2 2\theta_{14} \sim 0.3$. Experiment Neutrino-4 uses a much wider range of distances than STEREO and PROSPECT (FIG. 8 right) [12], which allows observing the process of oscillations directly in measurements using the method of coherent summation of measurement results with the same phase.

The talk [23] reports that the new results of the STEREO experiment exclude the 1σ region of the Neutrino-4 experiment with a reliability >3σ. We believe that the STEREO, PROSPECT experiments, should present the data in the form of the L/E dependence to compare their own results with the results of the Neutrino-4 experiment correctly. Only then the closing of the result of the Neutrino-4 experiment can be discussed.

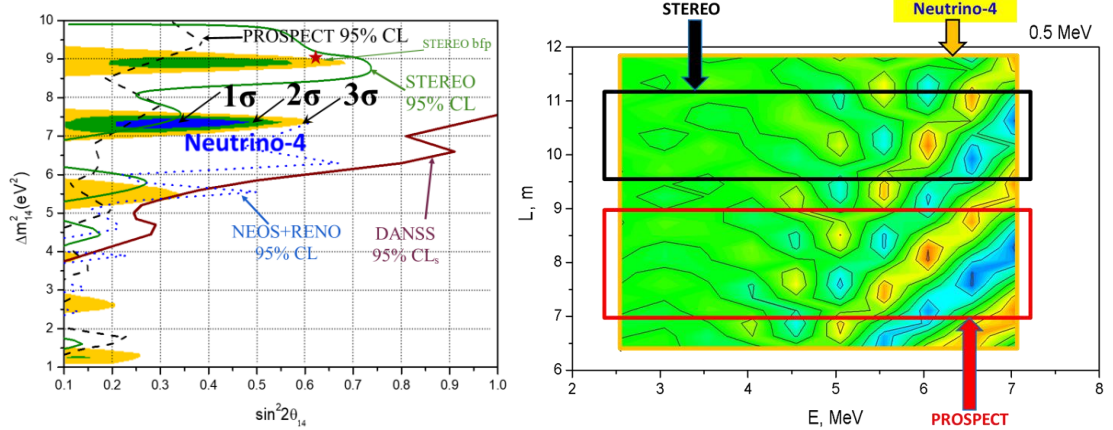

FIG. 8 Left - comparison of the sensitivity of the experiments in $0.1 < \sin^2 2\theta_{14} < 1$ and $1 < \Delta m^2 < 10$ eV$^2$ area: Neutrino-4, STEREO, PROSPECT, DANSS and NEOS. Right – comparison of regions (L, E) in measurements of the experiments: Neutrino-4, STEREO and PROSPECT.

### 5. Results of experiments KATRIN and GERDA

Now we should discuss the upper limit on $\Delta m_{14}^2$ from the results of the KATRIN and GERDA experiments. In FIG. 9, the region of sterile neutrino parameters is highlighted in blue, and this region is determined by the Troitsk, KATRIN, BEST, and DANSS experiments, and the result of the Neutrino-4 experiment is located within that region: $\Delta m_{14}^2 = (7.30 \pm 0.13_{st} \pm 1.16_{syst})$eV$^2$ $\sin^2 2\theta_{14} = 0.36 \pm 0.12$. The red ellipse indicates the 95% confidence level in the Neutrino-4 experiment. The result of the KATRIN experiment [24,25] does not exclude the region of Neutrino-4 $\sin^2 2\theta_{14} \leq 0.4$ [24]. Exclusion contours for the oscillation parameters in FIG. 9, including the limitations related to the consequences of the results of experiments on neutrinoless double beta decay, are taken from [24].

The GERDA experiment [26] requires special attention, since it is aimed at finding the mass of Majorana-type neutrinos. Currently, the majorana mass limit obtained in the GERDA experiment for the normal mass hierarchy is 1.6 standard deviation less than the majorana mass prediction obtained in the Neutrino 4 experiment. If in the future the Majorana mass limit of the experiment on double beta decay is lowered and the result of the Neutrino-4 experiment is confirmed, this will close the hypothesis light Majorana-neutrino.

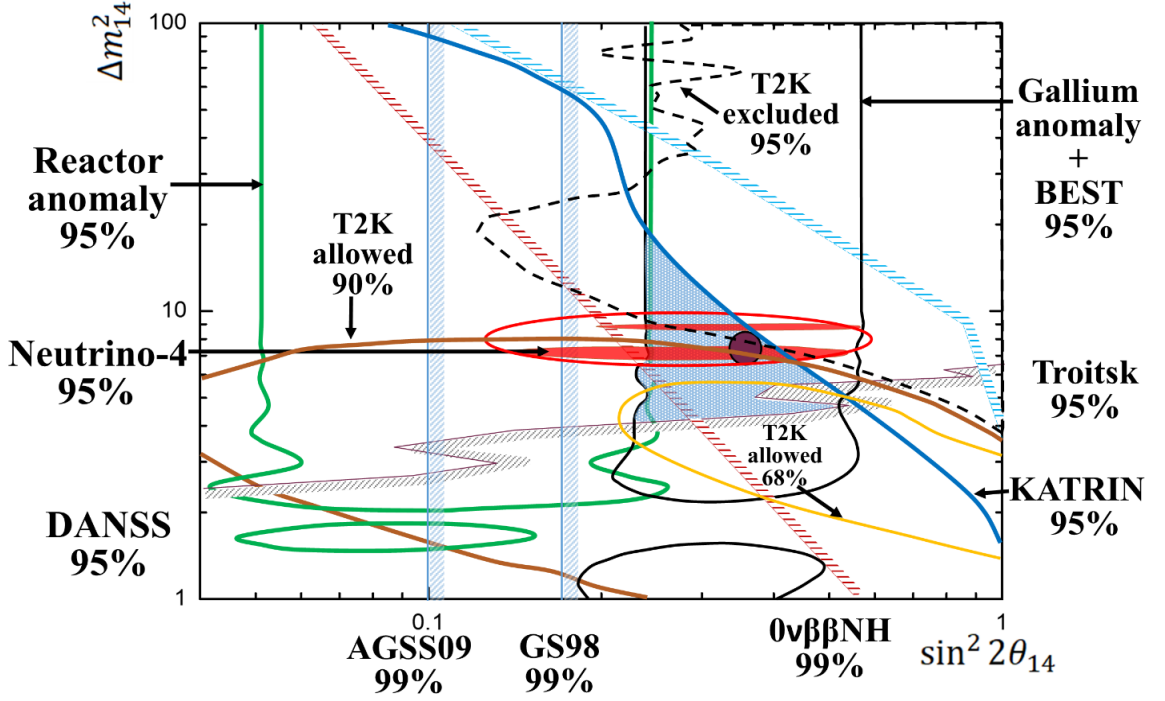

FIG. 9 The comparison of the results of the Neutrino-4 experiment with the results of the KATRIN and GERDA experiments. Exclusion contours are from [24] and [27].

**6. Ratios of different probabilities of oscillation processes in the 3 + 1 neutrino model.**

Discussion of neutrino experiments at accelerators in the context of the "Neutrino-4" result requires a brief description of the 3 + 1 neutrino model.

$$\begin{bmatrix} \nu_e \\ \nu_\mu \\ \nu_\tau \\ \nu_s \end{bmatrix} = \begin{bmatrix} U_{e1} & U_{e2} & U_{e3} & U_{e4} \\ U_{\mu 1} & U_{\mu 2} & U_{\mu 3} & U_{\mu 4} \\ U_{\tau 1} & U_{\tau 2} & U_{\tau 3} & U_{\tau 4} \\ U_{s1} & U_{s2} & U_{s3} & U_{s4} \end{bmatrix} \begin{bmatrix} \nu_1 \\ \nu_2 \\ \nu_3 \\ \nu_4 \end{bmatrix} \quad (4)$$

$$|U_{e4}|^2 = \sin^2(\theta_{14})$$
$$|U_{\mu 4}|^2 = \sin^2(\theta_{24}) \cdot \cos^2(\theta_{14}) \quad (5)$$
$$|U_{\tau 4}|^2 = \sin^2(\theta_{34}) \cdot \cos^2(\theta_{24}) \cdot \cos^2(\theta_{14})$$

$$P_{\nu_e \nu_e} = 1 - 4|U_{e4}|^2(1 - |U_{e4}|^2)\sin^2\left(\frac{\Delta m_{14}^2 L}{4E_{\nu_e}}\right) = 1 - \sin^2 2\theta_{ee} \sin^2\left(\frac{\Delta m_{14}^2 L}{4E_{\nu_e}}\right) \quad (6)$$

$$P_{\nu_\mu \nu_\mu} = 1 - 4|U_{\mu 4}|^2\left(1 - |U_{\mu 4}|^2\right)\sin^2\left(\frac{\Delta m_{14}^2 L}{4E_{\nu_\mu}}\right) = 1 - \sin^2 2\theta_{\mu\mu} \sin^2\left(\frac{\Delta m_{14}^2 L}{4E_{\nu_\mu}}\right) \quad (7)$$

$$P_{\nu_\mu \nu_e} = 4|U_{e4}|^2|U_{\mu 4}|^2 \sin^2\left(\frac{\Delta m_{14}^2 L}{4E_{\nu_e}}\right) = \sin^2 2\theta_{\mu e} \sin^2\left(\frac{\Delta m_{14}^2 L}{4E_{\nu_e}}\right) \quad (8)$$

It can be seen that if the mixing angles are small, then the new matrix elements $U_{e4}, U_{\mu 4}, U_{\tau 4}$ can be determined by measuring the mixing amplitudes. The probabilities of various oscillation processes are listed in expressions 6 to 8. Equation 6 represents the probability of electron neutrino disappearance due to oscillations to a sterile state. Equation 7 represents the probability of a muon neutrino disappearing due to oscillations to a sterile state. Equation 8 represents the probability of oscillation of a muon neutrino to an electron neutrino through a sterile state. The amplitudes of oscillations in these processes are:

$$\sin^2 2\theta_{ee} \equiv \sin^2 2\theta_{14} \quad (9),$$

$$\sin^2 2\theta_{\mu\mu} = 4\sin^2\theta_{24}\cos^2\theta_{14}(1 - \sin^2\theta_{24}\cos^2\theta_{14}) \approx \sin^2 2\theta_{24} \quad (10),$$

$$\sin^2 2\theta_{\mu e} = 4\sin^2\theta_{14}\sin^2\theta_{24}\cos^2\theta_{14} \approx \frac{1}{4}\sin^2 2\theta_{14}\sin^2 2\theta_{24} \quad (11).$$

Relatively small values of the mixing angles make it possible to simplify expressions without significant loss of accuracy. New parameters in simplified expressions (the square of the sine of the mixing angles) can be directly measured in experiments. The listed above relations of the parameters make it possible to compare the results of different types of oscillatory experiments. In the 3 + 1 neutrino model with one sterile neutrino, the oscillation length must be the same for all processes and be determined by the parameter $\Delta m_{14}^2$. Also, the amplitudes of the electron and muon neutrino oscillations in the disappearance processes determine the amplitude of the appearance of electron neutrinos in the muon neutrino flux $\sin^2 2\theta_{\mu e} \approx \frac{1}{4}\sin^2 2\theta_{14}\sin^2 2\theta_{24}$. This expression is very important for checking the validity of the 3 + 1 neutrino model. It has a fairly simple interpretation. The appearance of electron neutrinos in a flux of muon neutrinos is a second-order process, which can be considered as an oscillation of a muon neutrino into a sterile state and the subsequent oscillation of a sterile neutrino into an electron neutrino. Experiments in which effects are observed that indicate a transition to a sterile state in a flux of electron neutrinos are Neutrino-4, RAA and GA. Effects that can be interpreted as oscillation to a sterile state in a flux of muon neutrinos were observed in the IceCube[28,29] experiment, unfortunately, at a relatively low confidence level. Rather strict restrictions on $\sin^2 2\theta_{24}$ are presented in the work of [30] on the basis of all known works on measuring the disappearance effect in the muon sector. And finally, the experiments that claimed to observe the appearance of electron neutrinos in the flux of muon neutrinos are MiniBooNE[3] and LSND [2].

### 7. Comparison of Neutrino-4 results with IceCube, MiniBooNE and LSND results

To check the relation $\sin^2 2\theta_{\mu e} \approx \frac{1}{4}\sin^2 2\theta_{14} \sin^2 2\theta_{24}$ it is necessary, in addition to the data for $\sin^2 2\theta_{14}$, to have information about $\sin^2 2\theta_{24}$. The detailed analysis of the experimental data of the parameter $\sin^2 2\theta_{24}$ is shown in FIG. 10 from the work [28] about the IceCube experiment, where, in addition to the results of the experiment itself, limitations from a number of other experiments are presented. The red line in this figure is drawn by us as the envelope curve of the limits of all experiments at 99% CL. It excludes the best fit, but does not disprove the result of the IceCube experiment due to the large range of experimental uncertainties. These limits are quite conservative. Combining the results of two experiments with the same systematics cannot give an increase in the confidence level. In [30], a constraint obtained as a result of a joint analysis of the experiments shown in the figure is given, but, considering the combination of constraints, we will use a more conservative approach.

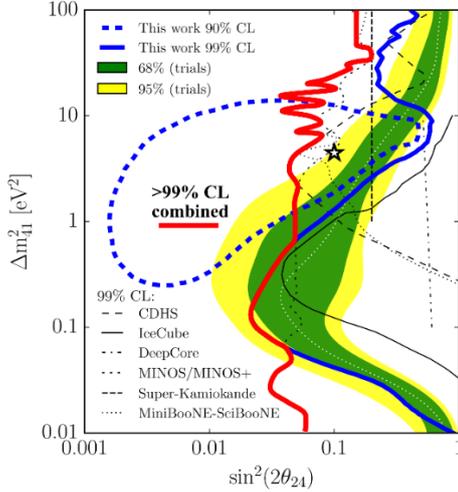

FIG. 10. The results of the IceCube experiment from [28], where, in addition to the results of the experiment itself, limitations from a number of other experiments are presented. Star is the IceCube best fit point.

For estimation of limitations on $\sin^2 2\theta_{\mu e} \approx \frac{1}{4}\sin^2 2\theta_{14} \sin^2 2\theta_{24}$ limitations from above on $\sin^2 2\theta_{24}$ for $\Delta m_{14}^2 = 1 \div 10 \text{ eV}^2$ mentioned earlier were used, and limitations from the Neutrino-4 on $\sin^2 2\theta_{14}$ for the same $\Delta m_{14}^2$ range were used. Limitation of product $\frac{1}{4}\sin^2 2\theta_{14} \sin^2 2\theta_{24}$ is presented if FIG. 11 on a linear scale (red line – limitations with CL 99%, red shaded area is allowed region). There are reslults from MiniBooNE - green and violet areas, LSND – gray line, KARMEN2 at 90%CL (data taken from [3]) and MicroBooNE with 3σ CL – blue line ([31]). The shaded area bounded by the red line and limits at $\Delta m_{14}^2$ includes the contours of the regions with ~3σ confidence obtained in the LSND and MiniBooNE experiments and do not contradict the current MicroBooNE result at the same confidence level. Thus, we can conclude that there are intersections at the level of contours with a confidence of 3σ, and thus, the possibility of reconciling the experimental results of the experiments: Neutrino-4, BEST and GA, MiniBooNE, LSND, IceCube within the framework of the 3+1 neutrino model is not completely closed.

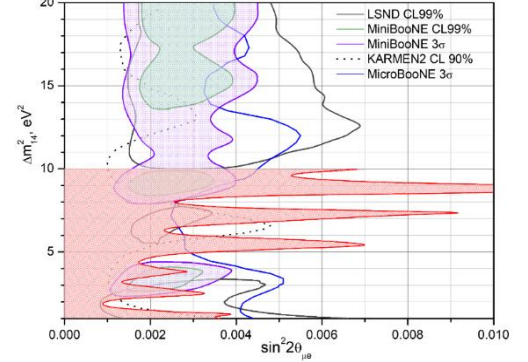

FIG. 11. A limit on $\sin^2 2\theta_{\mu e}$ (red line) as a result of the transferring restrictions on $\sin^2 2\theta_{24}$, using the relation $\sin^2 2\theta_{\mu e} \approx \frac{1}{4}\sin^2 2\theta_{14} \sin^2 2\theta_{24}$ and the value $\sin^2 2\theta_{14} = 0.36 \pm 0.23$ from the Neutrino-4 experiment. Red shading is the area of limitations of $\Delta m_{14}^2$ at the ~95% level from the Neutrino-4 experiment, taking into account the systematic uncertainties. Figure for comparison is taken from [3] and [31].

Comparison of the experimental data of MiniBooNE, LSND and Neutrino4 is shown in Fig. 12.

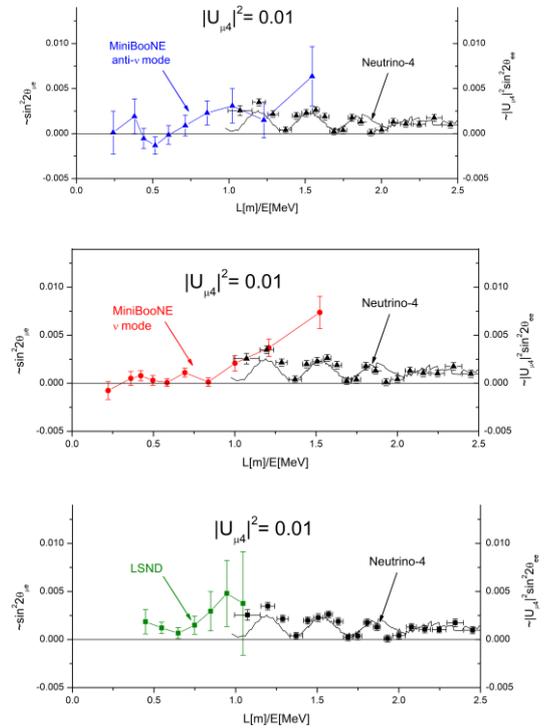

FIG.12. Comparison of the results of the Neutrino-4 experiment with the results of the MiniBooNE and LSND experiments [3] under the assumption that $\sin^2 2\theta_{24} \sim 2 \cdot 10^{-2}$ with the experimental value of $\sin^2 2\theta_{14} = 0.36 \pm 0.12$.

Unfortunately, the experimental accuracy at L/E <1 is still insufficient to observe an oscillatory dependence similar to the oscillatory dependence observed in the Neutrino-4 experiment.

## 8. Prediction of the effective mass of an electron neutrino obtained from the results of the Neutrino-4 experiment and comparison with the results of measuring the neutrino mass in the KATRIN experiment

The oscillation parameters obtained in the Neutrino-4 experiment can be used to estimate the effective mass of an electron neutrino in the 3+1 neutrino model using the well-known relation [1,32]:

$$m_{\nu_e}^{eff} = \sqrt{\sum m_i^2 |U_{ei}|^2}; \quad \sin^2 2\theta_{14} = 4|U_{14}|^2;$$

The sum of the neutrino masses $\sum m_\nu = m_1 + m_2 + m_3$ is limited by the results of cosmological studies by the value $0.087 \div 0.54$ eV at 95% CL (data taken from table 26.2 in [1] ). Therefore, in the limit $m_1^2, m_2^2, m_3^2 \ll m_4^2$, the effective mass of the electron neutrino can be calculated using the equation:

$$m_{\nu_e}^{eff} \approx \sqrt{m_4^2 |U_{e4}|^2} \approx \frac{1}{2}\sqrt{m_4^2 \sin^2 2\theta_{14}}.$$

The sterile neutrino mass can be estimated as: $m_4 = (2.70 \pm 0.22)$ eV. Using the value $\sin^2 2\theta_{14} \approx 0.36 \pm 0.12$ and $\Delta m_{14}^2 \approx (7.3 \pm 1.17)$ eV$^2$, one can estimate the mass of the electron neutrino:

$m_{4\nu_e} = (0.86 \pm 0.21)$ eV

The KATRIN experiment is aimed at direct measurement of the neutrino mass using the process T $\rightarrow$ He$_3 + e^- + \bar{\nu}$, near the 18.6 keV threshold. The square of the effective mass of the electron neutrino obtained by the KATRIN collaboration is $m_{3\nu_e}^2 = (0.26 \pm 0.34)$ eV$^2$, and the upper limit is m$_{3\nu_e}^{eff} \leq 0.8$ eV CL 90% [25]. This result was obtained within the framework of a 3-neutrino model with a 3x3 unitary PMNS matrix, therefore this result cannot be directly compared with the result obtained by analyzing the Neutrino-4 data.

Estimates in the framework of the 3 + 1 model under the assumption $m_1^2, m_2^2, m_3^2 \ll m_4^2$ are presented in Table 1 in the column Neutrino-4, and in the column KATRIN the results of the analysis are presented within the framework of 3 neutrinos hypothesis, without the sterile state. In principle, the KATRIN experimental data can be used for analysis within the 3+1 model using the parameters of the fourth neutrino obtained in the Neutrino-4 experiment $\sin^2 2\theta_{14} \approx 0.36 \pm 0.12$ and $m_4^2 \approx 7.3$ eV$^2$, to obtain an estimate for the masses $m_{1,2,3}^2$.

**Table 1** Estimation of the effective mass of an electron neutrino

|  | Neutrino-4 | KATRIN |
|---|---|---|
| Effective mass and mass squared ( m$_{\nu_e}^{eff}$, (m$_{\nu_e}^{eff}$)$^2$ ) | m$_{4\nu_e}^{eff} = (0.86 \pm 0.21)$ eV  (m$_{4\nu_e}^{eff})^2 = (0.73 \pm 0.29)$ eV$^2$ | m$_{3\nu_e}^{eff} < 0.8$ eV (90%)  (m$_{3\nu_e}^{eff})^2 = (0.26 \pm 0.34)$ eV$^2$ |

We calculated the contribution from a sterile neutrino to the differential spectrum of beta decay of tritium using the parameters $\sin^2 2\theta_{14} \approx 0.36 \pm 0.12$ and $m_4^2 \approx 7.3$ eV$^2$. This contribution is extremely small, so a logarithmic scale is used to demonstrate it in the figure 13 below. It is this small correction that needs to be introduced into the experimental data of the KATRIN experiment in order to derive an estimate for the squared masses $m_{1,2,3}^2$

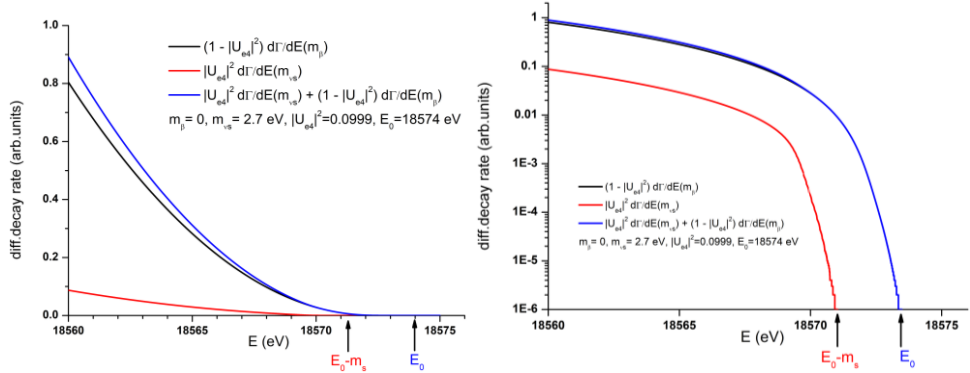

FIG.13. The calculation of the differential spectrum of electrons in the tritium beta decay taking into account the sterile neutrino with the parameters $\sin^2 2\theta_{14} \approx 0.33 \pm 0.07 (4.9\sigma)$ and m$_4^2 \approx 7.3$ eV$^2$, the red line shows the contribution of sterile neutrinos.

## 9. Comparison with neutrino mass constraints from experiments searching for double neutrinoless beta decay

Let us now consider the estimate for the effective Majorana mass of the electron neutrino. The effective Majorana mass of an electron neutrino is a parameter in the analysis of experiments on double β-decay and is determined by the ratio:

$m_{ee} = |\sum_i m_i U_{ei}^2| =$

$\begin{cases} \left|m_0 c_{12}^2 c_{13}^2 + \sqrt{\Delta m_{21}^2 + m_0^2} s_{12}^2 c_{13}^2 e^{2i(\eta_2 - \eta_1)} + \sqrt{\Delta m_{32}^2 + \Delta m_{21}^2 + m_0^2} s_{13}^2 e^{-2i(\delta_{CP} + \eta_1)}\right| \text{ in NO,} \\ \left|m_0 s_{13}^2 + \sqrt{m_0^2 - \Delta m_{32}^2} s_{12}^2 c_{13}^2 e^{2i(\eta_2 + \delta_{CP})} + \sqrt{m_0^2 - \Delta m_{32}^2 - \Delta m_{21}^2} c_{12}^2 c_{13}^2 e^{2i(\delta_{CP} + \eta_1)}\right| \text{ in IO} \end{cases}$

depending on the direct or inverse hierarchy of masses. However, for large values of $\Delta m_{14}^2$ this expression can be simplified to $m(0\nu\beta\beta) \approx m_4 U_{14}^2$. The quantitative value is $m(0\nu\beta\beta) = (0.27 \pm 0.12)$ eV. The strictest limitation on the Majorana neutrino mass is the result of the GERDA experiment [26]. This experiment aims to measure the half-life of the isotope, which is a function of the mass of the Majorana neutrino in the model with neutrinoless double beta decay. The lower limit with the 90% CL on the half-life of the isotope $^{76}$Ge is $T_{1/2}^{0\nu} > 1.8 \times 10^{26}$ years, which corresponds to the lower limit of the Majorana neutrino mass: $m_{\beta\beta} < [79 - 180]$ meV at 90% CL.

The value obtained with the oscillation parameters from the Neutrino-4 experiment is $m(0\nu\beta\beta) = (0.27 \pm 0.12)$ eV. The difference between the smallest value of the mass limit from the GERDA experiment and the estimate obtained from the Neutrino-4 results differ by 3 times. If in the future the limit of the Majorana mass from the experiment on double beta decay is lowered, and Neutrino-4 and GA with BEST results is confirmed, then this will close the hypothesis that the neutrino is a particle of the Majorana type.

## 10. Cosmological limitations and the results of Neutrino-4

The sum of the neutrino masses $\sum m_\nu = m_1 + m_2 + m_3$ is limited by the results of cosmological studies by the value $0.087 \div 0.54$ eV at 95% CL (data taken from table 26.2 in [1]). The introduction of one more neutrino with a mass of 2.7 eV causes obvious contradictions with these restrictions. In addition, there is an estimate [18] on the contribution of sterile neutrinos to the energy density of the Universe: $\Omega_{\nu_s} \simeq 0.2 \cdot \left(\frac{\sin 2\theta_\alpha}{10^{-4}}\right)^2 \cdot \left(\frac{m_\nu}{1 \text{keV}}\right)^2$. Sterile neutrinos with mass $m_\nu > 1$ keV and small mixing angle $\theta \leq 10^{-4}$ can be discussed as candidates for dark matter particles. More detailed consideration is in [33].

## 11. PMNS matrix in the 3+1 neutrino model

The material collected in the course of this analysis allows us to proceed to an attempt to construct the PMNS matrix in the $3 + 1$ neutrino model without consideration CP-violation phase, so it is matrix of module of the PMNS matrix. For this, the following values can be used: 1) $\sin^2 2\theta_{14} \approx 0.36 \pm 0.12$ from experiments Neutrino-4, 2) $\sin^2 2\theta_{24} = 0.027 \pm 0.014$ value is taken as the median value in the interval for $\sin^2 2\theta_{24}$ between the limit obtained by the IceCube experiment (blue line in FIG. 10 or [28] Figure 4) and the strongest limit in this region (short dotted line [28] FIG. 4 or red line in FIG. 10) for $\Delta m_{14}^2 = 7.3$ eV$^2$, 3) $\sin^2 2\theta_{34} \leq 0.16$ (limit obtained from Figure 19 from [29]), 3) $\sin^2 2\theta_{34} \leq 0.20$ (the limitation is derived from FIG. 19 of the work [29]). Then PMNS matrix in the 3-neutrino neutrino model:

$$U_{PMNS}^{(3)} = \begin{pmatrix} 0.824^{+0.007}_{-0.008} & 0.547^{+0.011}_{-0.011} & 0.147^{+0.003}_{-0.003} \\ 0.409^{+0.036}_{-0.060} & 0.634^{+0.022}_{-0.065} & 0.657^{+0.044}_{-0.014} \\ 0.392^{+0.025}_{-0.048} & 0.547^{+0.056}_{-0.028} & 0.740^{+0.012}_{-0.048} \end{pmatrix}$$

can be modified into $3 + 1$ model matrix.

$$U_{PMNS}^{(3+1)} = \begin{pmatrix} 0.778^{+0.024}_{-0.023} & 0.522^{+0.022}_{-0.021} & 0.147^{+0.005}_{-0.005} & 0.316^{+0.059}_{-0.059} \\ 0.472^{+0.030}_{-0.036} & 0.457^{+0.029}_{-0.037} & 0.709^{+0.016}_{-0.025} & 0.078^{+0.022}_{-0.022} \\ 0.272 \ldots 0.330 & 0.678 \ldots 0.708 & 0.614 \ldots 0.657 & 0 \ldots 0.211 \\ 0.212 \ldots 0.282 & 0.047 \ldots 0.208 & 0.108 \ldots 0.256 & 0.892 \ldots 0.946 \end{pmatrix}$$

Using this $3 + 1$ model matrix, the effective neutrino masses can be calculated by taking the most strict cosmological estimate for $\sum m_i = 0.11$ eV, and $m_1 = m_2 = m_3 = 0.037 \pm 0.040$ eV. The limit on the sum of masses is taken from the PDG, the error is based on total uncertainty. And finally, the value of the sterile neutrino mass: $m_4 = (2.70 \pm 0.22)$ eV.

$$m_1 = (0.037 \pm 0.04) \text{ эB} \quad m_{\nu_e}^{\text{eff}} = (0.86 \pm 0.21) \text{ эB} \quad \left(m_{\nu_e}^{\text{eff}}\right)^2 = (0.73 \pm 0.36) \text{ эB}^2$$

$$m_2 = (0.037 \pm 0.04) \text{ эB} \quad m_{\nu_\mu}^{\text{eff}} = (0.21 \pm 0.07) \text{ эB} \quad \left(m_{\nu_\mu}^{\text{eff}}\right)^2 = (0.05 \pm 0.03) \text{ эB}^2$$

$$m_3 = (0.037 \pm 0.04) \text{ эB} \quad m_{\nu_\tau}^{\text{eff}} = (0.037 \div 0.53) \text{ эB} \quad \left(m_{\nu_\tau}^{\text{eff}}\right)^2 = (0.0014 \div 0.27) \text{ эB}^2$$

$$m_4 = (2.70 \pm 0.22) \text{ эB} \quad m_{\nu_e}^{\text{eff}} = (2.51 \div 2.55) \text{ эB} \quad \left(m_{\nu_s}^{\text{eff}}\right)^2 = (6.26 \div 6.53) \text{ эB}^2$$

Below we present a mixing scheme for the active neutrino flavors and the sterile neutrino for the normal and inverse mass hierarchy (FIG.14). However, it should be noted that in both cases the main effect of the mass hierarchy is due to the difference in the masses of the sterile neutrino and the SM neutrinos.

It can be seen that the sterile neutrino provides the effective masses of all neutrinos, while the masses of active neutrinos $m_1, m_2, m_3$ are small and close to each other.

It can be seen the sterile neutrino provides the electron neutrino effective mass, under the current cosmological constraints on $m_1, m_2, m_3$.

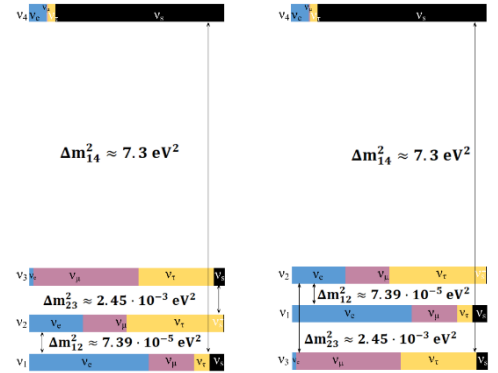

FIG. 14. Scheme of mixing of active neutrinos and the sterile neutrino for normal and inverse mass hierarchies.

## 12. Conclusions of the presented analysis

1. The results of direct experiments on the search for sterile neutrinos - Neutrino-4 and BEST with GA indicate the existence of a sterile neutrino with oscillation parameters: $\Delta m_{14}^2 = 7.30 \pm 0.13_{st} \pm 1.16_{syst}$ $\sin^2 2\theta_{14} = 0.35 \pm 0.07 (5.0\sigma)$. Values of the parameters corresponding to the best fit are $\Delta m_{14}^2 = 7.3$ эВ$^2$, $\sin^2 2\theta_{14} = 0.38$ and confidence level for the 4ν oscillation hypothesis is 5.8σ.
2. The analysis of the direct experimental data on the search for sterile neutrinos in accelerator experiments indicates the possibility to achieve the agreement of the results of experiments within the overlaping of contours with a confidence level of 2σ: Neutrino-4, BEST+GA, MiniBooNE, LSND, IceCube within the 3+1 neutrino model and at the current limits of experimental accuracy.
3. The range of values of the effect of neutrino appearance in the MiniBooNE, LSND experiments is not excluded, but is limited by the data on the disappearance of muon neutrinos using $\sin^2 2\theta_{14} = 0.36 \pm 0.12$ from experiments Neutrino-4 and data on $\Delta m_{14}^2 = 7.30 \pm 0.13_{st} \pm 1.16_{syst}$ from the Neutrino-4 experiment.
4. Parameters of sterile neutrino $m_4 = (2.70 \pm 0.22)$eV and $\sin^2 2\theta_{14} = 0.36 \pm 0.12$ allow us to estimate the effective mass of an electron neutrino: $m_{4\nu_e}^{\text{eff}} = (0.86 \pm 0.21)$eV. The KATRIN experiment does not observe a sterile neutrino and makes a limitation on the neutrino mass in the 3 neutrino model $m_{3\nu_e}^{\text{eff}} < 0.8$ eV (90%) therefore direct comparison of the results is not correct.
5. The parameters of the fourth neutrino obtained in the Neutrino-4 experiment should be used in the KATRIN experiment as additional parameters for determining $m_{3\nu_e}^{\text{eff}}$.
6. The value of the Majorana mass, calculated with the parameters of the oscillations obtained in the Neutrino-4, is $m(0\nu\beta\beta) = (0.27 \pm 0.12)$ eV, which is three times higher than the limit declared by the GERDA experiment. If in the future the limit of the Majorana mass in the experiment on double neutrinoless beta decay is lowered and Neutrino-4 and BEST with GA results is conferred, then this will close the hypothesis that the neutrino is a particle of the Majorana type.
7. STEREO, PROSPECT experiments for a correct comparison of their own results with the results of the Neutrino-4 experiment should present the data in the form of the L/E dependence. Only then we can discuss the closure or confirmation of the result of the Neutrino-4 experiment.

## 13. Prospects for the development of experiments measuring sterile neutrino parameters

First of all, practically all the experiments discussed above will continue. Here we would like to draw your attention to experiments in which the measurement accuracy will be significantly increased.
1. At the ICARUS facility at FermiLab, Carlo Rubbia plans high-precision experiments [34]. These experiments will primarily focus on the following four problems.
   Observation of the disappearance of muon neutrinos.
   Detecting the disappearance of electron neutrinos.
   Detection of the appearance of electron neutrino signals in a muon neutrino beam.
   Confirmation of effective masses of sterile neutrinos.
Thus, this is a complete plan for solving the sterile neutrino problem. Work in progress.
2. The Neutrino-4 collaboration plans to improve the existing experimental setup, as well as create a second neutrino laboratory at the SM-3 reactor, which will be equipped with a detector with three times higher sensitivity [35]. Work in progress.

### Acknowledgements

The work was supported by the Russian Science Foundation under Contract No. 20-12-00079.
The authors are grateful to the colleagues of PNPI NRC KI and INR RAS for useful discussions at the seminars. The Neutrino-4 collaboration would like to thank Carlo Rubbia for drawing attention to the results of our experiment.